\documentclass[pra,
                preprint,
                amsmath,
                amssymb,
                showpacs,
                superscriptaddress]{revtex4}
\usepackage{graphicx} 
\usepackage{dcolumn}  
\usepackage{bm}       
\usepackage{amsfonts}
\usepackage{dsfont}
\usepackage{csquotes}
\usepackage{ulem}
\usepackage{color}

\usepackage{float}
\usepackage{physics}
\usepackage{soul,xcolor}
\setstcolor{red}

\begin{document}

\title{Probing the Effect of Molecular Structure Saddling on Ultrafast Charge Migration via Time-Resolved X-ray Diffraction}

\author{Sucharita Giri}
\affiliation{%
Department of Physics, Indian Institute of Technology Bombay,
            Powai, Mumbai 400076  India}
            
\author{Jean Christophe Tremblay}
\affiliation{%
Laboratoire de Physique et Chimie Th\'eoriques, 
CNRS-Universit\'e de Lorraine, UMR 7019, ICPM, 1 Bd Arago, 57070 Metz, France}

\author{Gopal Dixit}
\email[]{gdixit@phy.iitb.ac.in}
\affiliation{%
Department of Physics, Indian Institute of Technology Bombay,
            Powai, Mumbai 400076  India}

\date{\today}


\begin{abstract}
Metal-corroles are macrocycle organic molecules with numerous practical applications.
In particular, copper corroles  exhibit an interesting  saddled geometry, which has 
attracted significant  attention from theoreticians and experimentalists over the years.
The present work is dedicated  to understand the effect of structural saddling in a copper corrole 
on potential probe signals via  imaging ultrafast coherent electron dynamics. 
A linearly polarized pulse is used to trigger the electron dynamics
and time-resolved x-ray diffraction is employed to image the triggered dynamics. 
It is found that the  symmetry reduction in the 
time-resolved diffraction signals and electronic  flux densities  is a signature of the saddling in a copper corrole during ultrafast charge migration.
Moreover,  analysis of the electronic flux density  reveals that the diagonal nitrogen atoms 
mediate coherent charge migration between them via a central copper atom.
Correlation of the  flux densities and the diffraction signals indicates that 
the signature of  the charge migration is encoded in time-resolved diffraction signals.  
A comparison of the static diffraction signals of nonsaddled planar copper porphyrin and 
saddled nonplanar copper corrole in their ground states is made.
\end{abstract}

\maketitle 

\section{Introduction}
The chemistry of porphyrins and corroles is a never-ending field of exciting problems, 
for both fundamental and practical reasons~\cite{mingos2012molecular}. 
Corroles and porphyrins are macrocyclic organic molecules as they sustain a conjugation
channel bearing a large number (18) of $\pi$ electrons.
Due to a slightly smaller number of atoms,  
corroles are commonly known as reduced dimensional models for porphyrins~\cite{johnson1965corroles},
with whom they share similar electronic properties.
Over the years, many significant research works have been performed 
to understand ultrafast light-induced 
processes in porphyrins from a theoretical perspective~\cite{tremblay2021time, nam2020monitoring, koksal2017effect, barth2006periodic, barth2006unidirectional}.  
In contrast, corroles have not received similar attention in the context of light-induced ultrafast processes~\cite{lemon2020corrole}. 
The present work is a first step towards exploring ultrafast coherent electron dynamics in 
metal-coordinated corroles.  

\begin{figure}[h!]
\includegraphics[width = 0.9 \linewidth]{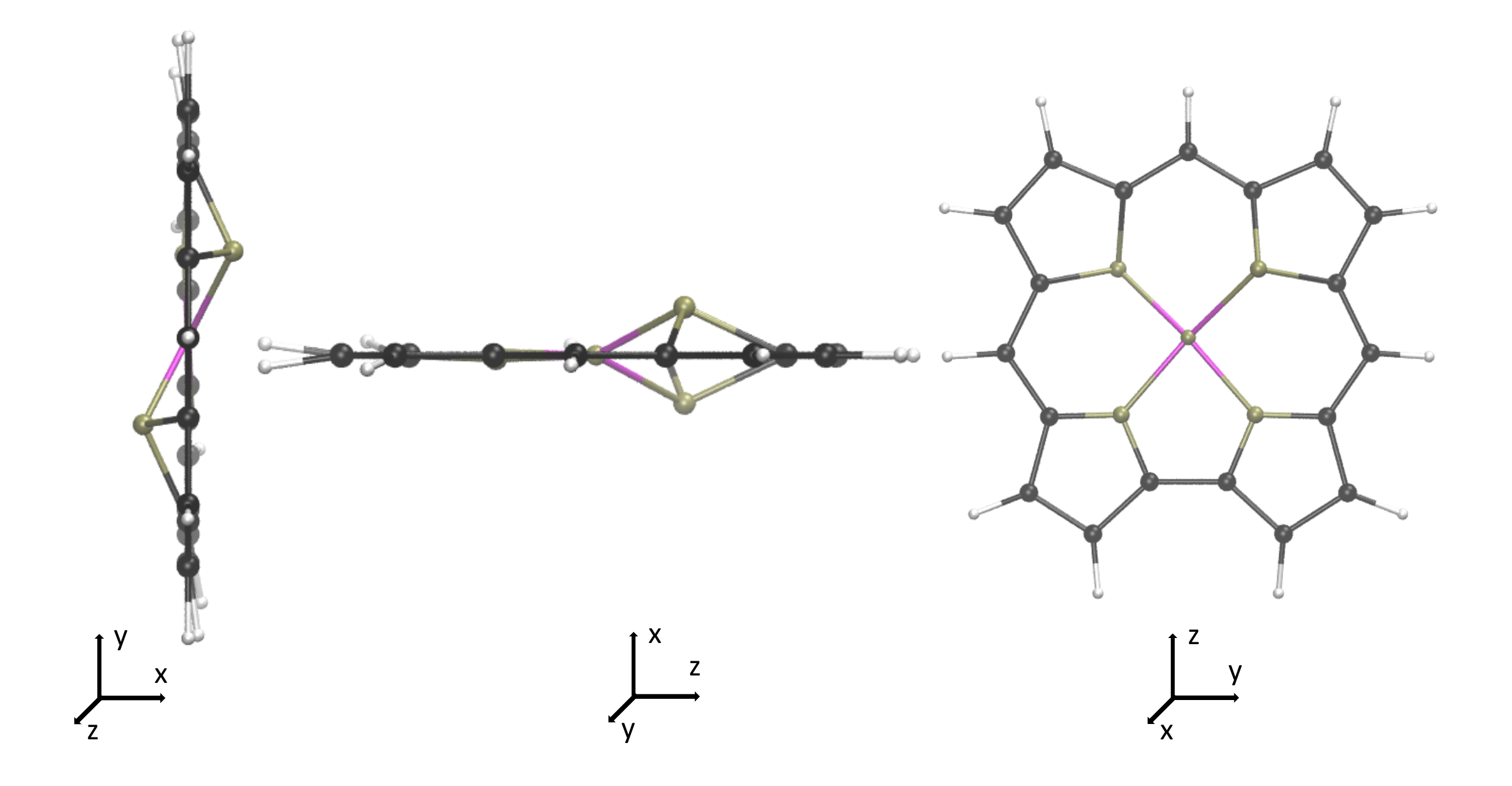}
\caption{Ball-stick representation of unsubstituted  copper corrole in different orientations, with 
the molecule  in the $yz$-plane.
Magenta, tan, black, and white spheres represent copper, nitrogen, carbon, and hydrogen atoms, respectively.} \label{fig01}
\end{figure}

Technological advancements in recent years have allowed  one to control the synthesis of different types of corroles with interesting coordination chemistry. 
The properties of the central metal in metal-corroles are always the focus of ample attention. 
Due to the intriguing coordination chemistry and its compatibility with a variety of transition metals~\cite{nardis2019metal}, 
metal-corroles have demonstrated potential for applications as 
photosensitizers~\cite{jiang2019corrole, mahammed2019corroles} and 
catalysts~\cite{dogutan2011electocatalytic, gross2000epoxidation, mahammed2003aerobic}, to name but a few.
Copper corrole is one of the first systems bearing a transition metal in which copper was experimentally inserted in the N$_4$ coordination core of the corrole ring. 
Unlike other metal-corrole complexes, 
copper corrole is found to have a saddling in its equilibrium structure (see Fig.~\ref{fig01}), which makes this molecule very interesting for different applications~\cite{ghosh2000electronic, luobeznova2004electronic, broring2007revisiting, pierloot2010copper}. 
A combined study of x-ray absorption spectroscopy and time-dependent density functional theory  (TDDFT) have been  performed 
to understand the electronic structure of copper corrole~\cite{lim2019x}. 
Moreover, the role of the  saddling on the electronic structure properties of low-lying electronic states in copper corrole
was explored by analyzing the results of planar and saddled geometries~\cite{pierloot2010copper}. 
Furthermore, static x-ray diffraction with  density functional analysis confirmed the  saddling 
feature of copper corroles~\cite{alemayehu2009copper, broring2007revisiting}. 
Several theoretical works  have been carried out to understand the saddling in copper corroles~\cite{ghosh2000electronic,luobeznova2004electronic, alemayehu2009copper, lim2019x}. 

{\it A priori}, it is not obvious how the  saddling in a molecular structure will affect the coherent electron dynamics 
on attosecond timescales, during which the effect of nuclear vibrations is insignificant.
Moreover, what would be the signature of saddling in any experimental probe signal, if any? 
The main aim of the present work is to address such crucial questions. 
In this work, we present time-resolved imaging of coherent electron dynamics in an unsubstituted copper corrole within a pump-probe configuration. 
An ultrashort pump pulse induces a coherent electron dynamics, which is imaged by time-resolved x-ray diffraction (TRXD) at various pump-probe delay times.  
The availability  of ultrashort, intense x-ray pulses from various  
free-electron laser sources around the globe~\cite{ishikawa2012, emma2}
has opened a new array of possibilities to extend  x-ray diffraction from static to time domain. 
Furthermore, there are successful reports of  x-ray pulse generation in the attosecond time domain~\cite{hartmann2018attosecond, duris2020tunable}.
TRXD is an emerging method to probe ultrafast processes in nature with atomic-scale spatial and temporal resolutions and  has triggered 
significant theoretical~\cite{topfer2021imaging, giri2021imaging, simmermacher2020time, hermann2020probing, simmermacher2017time, kowalewski2017monitoring, dixit2017time, slowik2014incoherent, bredtmann2014x, dixit2013jcp, dixit2013prl, dixit2012, simmermacher2019theory} and  
experimental~\cite{zhang2022ultrafast, yong2021ultrafast, yong2018determining, glownia2016self, minitti2015imaging} research on TRXD from different molecules in recent years. 
Additionally, we will analyze time-dependent electronic flux densities 
to understand the mechanistic details of the charge migration associated with the induced dynamics. The time-dependent  electronic flux densities provide additional information 
about the direction of the electron flow during charge migration. 

\section{Computational Methods}
The following expression of the differential scattering probability is used to simulate the 
TRXD corresponding to a coherent electron dynamics as~\cite{dixit2014theory}
\begin{equation}\label{eq1}
\frac{dP(\tau)}{d\Omega} =
\frac{dP_{e}}{d\Omega} \sum_{j} \left| 
\int \mathrm{d}\mathbf{r} ~\langle \Phi_{j}  | \hat{\rho}(\mathbf{r}) | \Psi(\tau) \rangle~ e^{-i \mathbf{Q} \cdot \mathbf{r}}\right|^{2}.
\end{equation}
Here, ${dP_{e}}/{d\Omega}$ is the Thomson scattering cross section of a free electron, 
$| \Psi(\tau) \rangle$ is an electronic wavepacket with $\tau$ as the pump-probe delay time, 
$| \Phi_{j}  \rangle$ is an eigenstate of the unperturbed electronic Hamiltonian of the system,
$\mathbf{Q}$ is the photon momentum transfer, and $\hat{\rho}(\mathbf{r})$ is the density operator. 
The above equation is valid for probe x-ray pulses shorter than the characteristic timescale of the dynamics. 

The one-electron density $\rho(\mathbf{r}, t)$ can be used to describe relevant  quantities corresponding to coherent many-electron dynamics.
It can be obtained as the expectation value of the one-electron density operator from 
the electronic wavepacket as 
\begin{eqnarray}\label{rho_op}
\hat{\rho} \left( \mathbf{r}\right) = \sum_{k}^{N} \delta\left( \mathbf{r} - \mathbf{r}_{k} \right).
\end{eqnarray}
To calculate $\rho(\mathbf{r}, t)$, the $N$-electron wavepacket $\Psi(\mathbf{r}^N, t)$ is obtained by solving the many-electron time-dependent Schr\"{o}dinger equation. 
In this work, the $N$-electron wavepacket is represented as a linear combination of the ground-state wave function, $\Phi_0(\mathbf{r}^N)$,
and of the lowest-lying many-body excited states of the system $\Phi_k(\mathbf{r}^N)$ as
\begin{equation}\label{tdci}
\Psi(\mathbf{r}^N,t) = \sum\limits_{k=0}^{N_{\rm states}} C_k(t) \Phi_k(\mathbf{r}^N).
\end{equation}
In the generic time-dependent configuration interaction formulation, the excited states are expressed  
as a linear combination of configuration state functions,
$\Phi_a^r(\mathbf{r}^N)$, obtained from a reference ground-state wave function.
In order to keep the computational cost low, only single excitations are taken into account, as 
\begin{equation}\label{cis}
\Phi_k(\mathbf{r}^N) = \sum\limits_{ar} D_{a,k}^r \Phi_a^r(\mathbf{r}^N).
\end{equation}
This provides a physically sound representation to the one-electron process investigated here.
In the single-excited configuration state functions, $a$ and $r$ represent the excitation from occupied molecular orbital $a$ to unoccupied orbital $r$. 
In the framework of the hybrid time-dependent density functional theory (TDDFT)/configurational interaction (CI)
methodology \cite{klinkusch2016,hermann2016tddft}, all expansion coefficients $\{D_{a,k}^r\}$ are obtained from linear-response TDDFT (LR-TDDFT). 
The LR-TDDFT calculation is performed once prior to all dynamical simulations using standard quantum chemistry packages.
The expansion coefficients of the excited states in Eq.\,\eqref{cis} are renormalized to form pseudo-CI eigenfunctions.
In the basis of pseudo-CI eigenfunctions chosen for the propagation, the Hamiltonian is diagonal to a good approximation.
The time evolution of the coefficients $C_k(t)$ in Eq.\,\eqref{tdci} is simulated by direct numerical integration
of the time-dependent Schr\"{o}dinger equation using a preconditioned adaptive step size Runge-Kutta algorithm \cite{04:TC:passrk}.
The electronic flux density is calculated using Eqs.~\eqref{tdci} and \eqref{cis}, as the expectation value of the operator of the following form
\begin{eqnarray}\label{j_op}
\hat{j} \left( \mathbf{r}\right) = \frac{1}{2}\sum_{k}^{N} \left(\delta\left( \mathbf{r} - \mathbf{r}_{k} \right) \hat{p}_{k} + \hat{p}^{\dagger}_{k}\delta\left( \mathbf{r} - \mathbf{r}_{k} \right)\right), 
\end{eqnarray}
where $\hat{p}_{k}=-i\vec{\nabla}_{k}$ represents the momentum operator. The open-source toolbox detCI@ORBKIT is used for post-processing the information obtained from the quantum chemistry package~\cite{hermann2016orbkit,pohl2017open, hermann2017open}.

In this work,  $N_\textrm{states}=80$ lowest-lying excited states below the ionization threshold are used to achieve the convergence of the 
excitation dynamics induced by the pump pulse. 
The states are computed using the CAM-B3LYP functional \cite{04:YTH:cam} and aug-cc-pVDZ basis sets \cite{dunning1989ccpvxz} on all atoms,
as implemented in Gaussian16~\cite{frisch2016gaussian}.
The frozen nuclei approximation is used throughout this work, as the motion of nuclei is typically much slower in comparison to the electronic motion.
The computation of the one-electron integrals from the many-body wave function is performed using the detCI@ORBKIT
toolbox~\cite{hermann2016orbkit,pohl2017open, hermann2017open}.
All many-electron dynamical simulations are performed using in-house codes \cite{klinkusch2011,04:TC:passrk,08:TKS,08:TS:gloct,10:TKKS}.
The integrals required to compute the TRXD signals and electronic flux densities are calculated using ORBKIT~\cite{OKgit}.

\section{Results and Discussion}

The molecular structure of the unsubstituted copper corrole 
is shown in Fig.~\ref{fig01} in ball-stick representation.
All axes and coordinates are described in the molecular frame of reference,
and it is assumed that the molecule is oriented in the $yz$-plane of the laboratory frame. 
From the projections in the two other planes, the saddling in copper corrole is evident, 
which makes this particular type of corrole  appealing for detailed investigations. 

\begin{figure}[h!]
\includegraphics[width=0.5\linewidth]{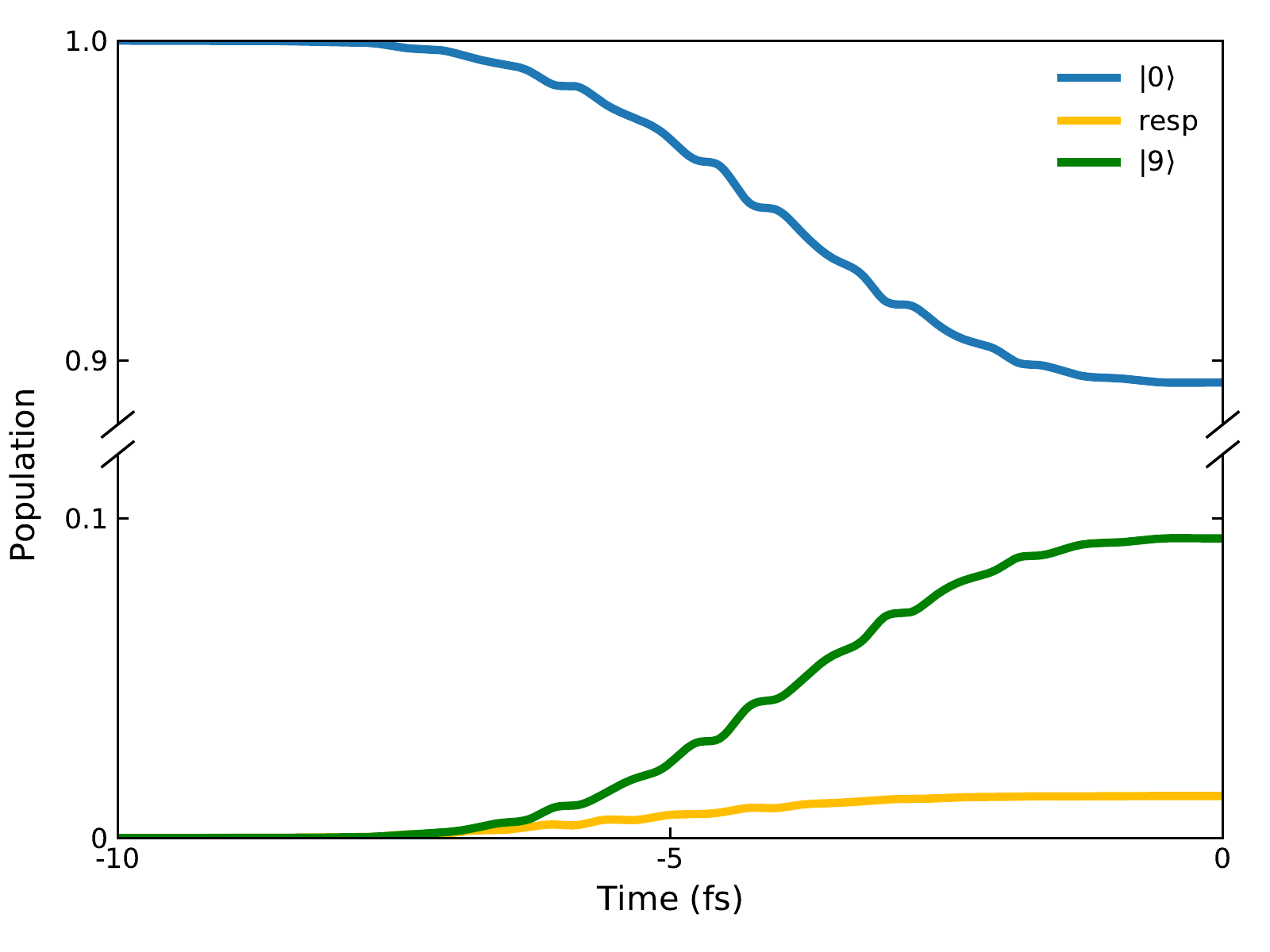}
\caption{Population dynamics of selected electronic states of copper corrole. 
 $\vert 0 \rangle$ represents the ground electronic state, and $\vert 9 \rangle$ represents 
the target electronic state with energy E$_{9}$ = 3.45 eV.
The orange line labeled ``resp'' represents the electronic response in the presence of the field, leading to the 
residual population of the rest of the electronic states after the excitation.
A cosine-squared linearly polarized pulse of 360 \,nm wavelength with peak intensities of 
3.5$\times$10$^{13}$ W/cm$^{2}$ is used to induce the population dynamics.  
The pulse is 10\,fs short and polarised along the $y$ axis.  
Time zero defines the onset of field-free electron dynamics.} \label{fig02}
\end{figure}

To trigger the electron dynamics, 
a 10 fs cosine-squared pulse linearly polarized along the $y$-axis is employed.
The wavelength and the peak intensity of the pump pulse are chosen to 360 \,nm  and 
3.5$\times$10$^{13}$ W/cm$^{2}$, respectively.
For an ionization potential of about 6.78\,eV, the Keldysh parameter is found to be 2.8, such that electron loss to the ionization channel via multiphoton processes is not dominant. 
It has been experimentally demonstrated that a coherent transfer of population from ground to excited states is possible with such intensity in molecular systems~\cite{prokhorenko2005coherent}.
The resulting population dynamics of the electronic states during the pump pulse is presented in Fig.~\ref{fig02}. 

Transfer of population from the electronic ground state to other excited states starts as the pump pulse interacts  and 
the electron dynamics in  copper corrole sets into motion.  
As evident from Fig.~\ref{fig02}, only a small amount of population around $9.4\%$ is transferred to the ninth  electronic excited state,
and most of the population remains in the ground state, i.e., $89\%$ at the end of the pulse. 
Small population transfer from the ground to excited states is common in various  excitation schemes during experiments~\cite{liu2018attosecond, glownia2016self}, 
as it can be achieved in a more controlled manner with lower field intensities.
The rest of the electronic populations, termed as ``resp'' in Fig.~\ref{fig02}, is distributed among other excited states with insignificant probability. 
The insignificant contributions from other states involved in the electronic response during the excitation can be neglected for further analysis.
The energy  difference between the ground  and the ninth excited states is $\Delta {\textrm{E}} = 3.45$\,eV. 
Focusing only on the dominantly populated state allows one to estimate a characteristic timescale for the electron dynamics as
$\textrm{T} = \hbar/ \Delta {\textrm{E}}$= 1.2\,fs, which is used for discussions.
Time zero in Fig.~\ref{fig02} represents the onset of field-free charge migration, i.e., after the pump pulse ended. 

\begin{figure}[h!]
\includegraphics[width=1.0\linewidth]{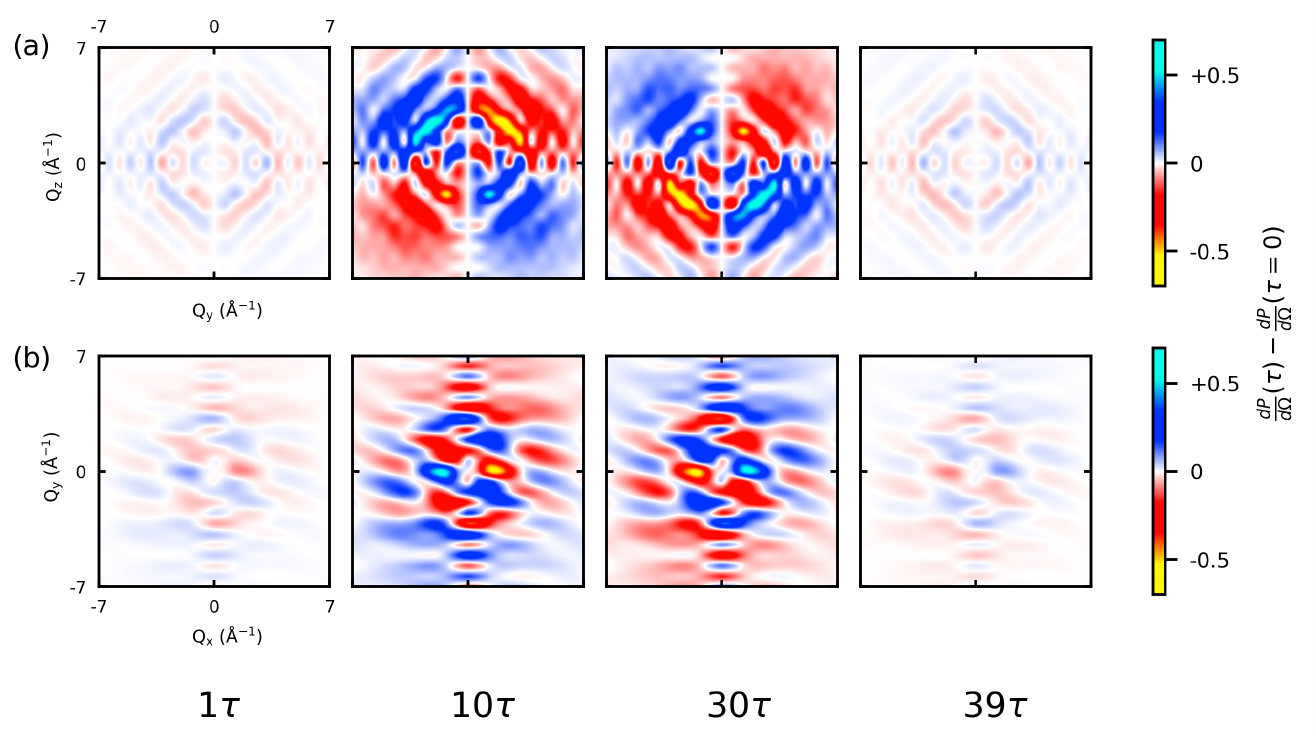}
\caption{Time-resolved difference diffraction signals for copper corrole in (a) $Q_{y}-Q_{z}$ and 
(b) $Q_{x}-Q_{y}$ planes at different pump-probe delay times during field-free charge migration.
Here,  $\tau = \frac{\textrm{T}}{40}$ is chosen with $\textrm{T} = 1.2$ fs as  the characteristic timescale of the electron dynamics.
The intensity of the diffraction patterns is presented in units of  ${dP_{e}}/{d \Omega}$.  
The time-independent diffraction signal at zero delay time is subtracted at all subsequent delay times.} \label{fig03}
\end{figure}

After triggering the coherent electron dynamics by a short, intense pump pulse, we employ
TRXD in a pump-probe configuration to probe the dynamics using a probe pulse much shorter than $\tau$ = 30 as.
Figure \ref{fig03} presents time-resolved diffraction signals at different pump-probe time delays during the field-free charge migration dynamics.
As the saddling is present in both the  $zx$ and $xy$ planes, while 
the molecule lies in the $yz$ plane without any saddling (see Fig.~\ref{fig01}),
the diffraction signals  are presented  in $Q_{y}-Q_{z}$ and $Q_{x}-Q_{y}$ planes, which  are  
used to emphasize the role of saddling during the electron dynamics.
For representation purposes, the total signal at the zero pump-probe delay time is subtracted from the total signals at different subsequent delay times.
Ground and the 10 lowest-lying excited states are used to simulate the TRXD signal, 
i.e., $j = [0,10]$ in Eq.\,\eqref{eq1}.

\begin{figure}[h!]
\includegraphics[width= 0.8\linewidth]{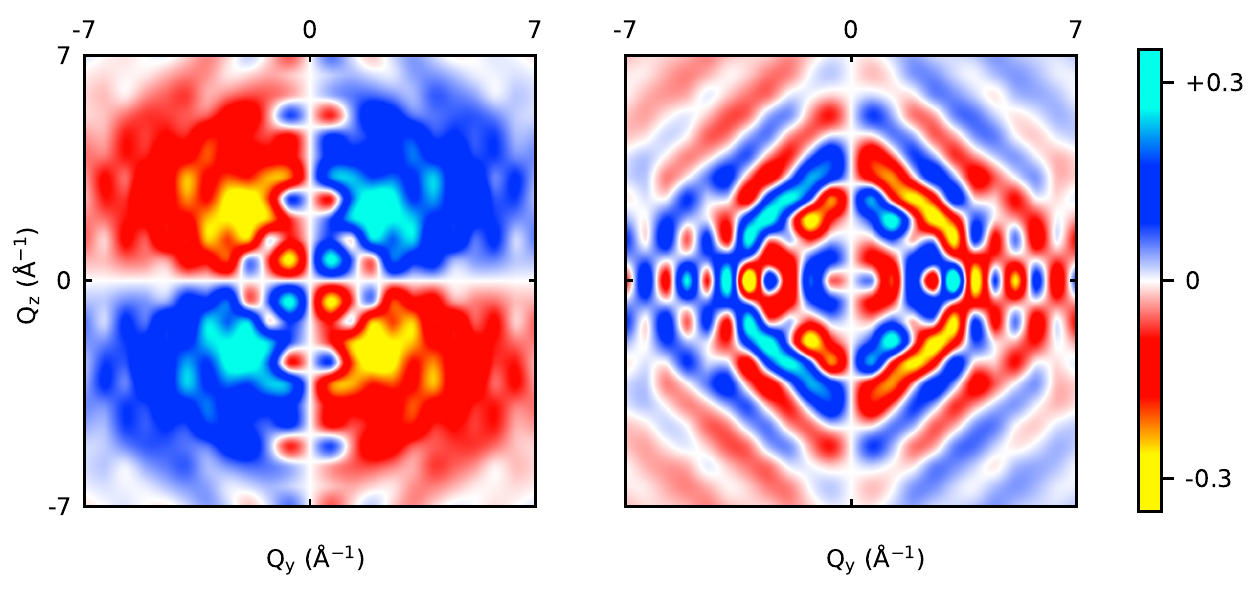}
\caption{Left panel: Real part;  Right panel: imaginary part of  $\mathcal{L}_{90}(\mathbf{Q})$ in  the 
$Q_{y}-Q_{z}$ plane.} \label{fig003}
\end{figure}

At a glance, it seems that the overall intensity variation of the diffraction signals in  the
$Q_{y}-Q_{z}$ [Fig.\,\ref{fig03}(a)]  and $Q_{x}-Q_{y}$  [Fig.\,\ref{fig03}(b)]   planes is approximately  
the same at all delay times, up to the phase and the signal magnitude. 
The global maxima and smaller local features migrate up and down in the $Q_{y}-Q_{z}$ plane.
In the upper panels, the signal is found to be antisymmetric with respect to reflection about the $Q_{y} = 0$ line.
At the beginning of the charge migration, i.e., at delay time $1\tau$, the signal also exhibits reflection symmetry with respect to the $Q_{z} = 0$ line around low $\mathbf{Q}$ region.  
This feature in the signal is also true around the characteristic timescale $\textrm{T} = 1.2$ fs $= 40 \tau$, 
which is clearly visible from the signal at $39\tau$ [the last figure of  Fig.\,\ref{fig03}(a)].
Note that the total signal, which is not the difference signal,  is strictly anti-symmetric at $\textrm{T} = 0$ and $40\tau$, and the snapshots at $\tau$ and $39\tau$
are chosen at the onset of symmetry reduction.
This scenario changes significantly at the two intermediate delay times, $10\tau$ and $30\tau$. 
The intensity of the signal exhibits an extremum close to the center, with local maxima at around values of $\pm 1$ in $Q_{y}$ and $Q_{z}$.
The extremum is initially more intense and diffused in the upper half of the plane, i.e., at positive values along the $Q_{z}$ axis at delay time $10\tau$.
The symmetry reduction can be understood using a simplified model of the charge migration as a superposition of $|0\rangle$ and $|9\rangle$ electronic states.
As derived from Eq.\,\eqref{eq1} (see also SI of Ref.\,[\citenum{hermann2020probing}]), the time-dependent part of the
TRXD signal in such superposition state reads
\begin{equation}\label{tls}
\frac{dP(\tau)}{d\Omega} \propto  \cos(\omega_{09}\tau)\,\text{Re}\big[{\mathcal L}_{09}(\mathbf{Q})\big]
-\sin(\omega_{09}\tau)\,\text{Im}\big[ {\mathcal L}_{09}(\mathbf{Q})\big],
\end{equation}
where $\omega_{09}=(E_9-E_0)/\hbar$ is the transition frequency between the $|0\rangle$ and $|9\rangle$  states, and
\begin{equation}
{\mathcal L}_{09}(\mathbf{Q}) = \sum_{j}~ \int \mathrm{d}\mathbf{r}  \int \mathrm{d}\mathbf{r}^{\prime} ~ \langle \Phi_{0}  | \hat{\rho}(\mathbf{r}) | \Phi_{j} \rangle \langle \Phi_{j}| \hat{\rho}(\mathbf{r}^{\prime})  | \Phi_9 \rangle~ e^{i \mathbf{Q} \cdot (\mathbf{r} -\mathbf{r}^{\prime})}.
\end{equation}
As evident from Fig.~\ref{fig003}, the real and imaginary parts of ${\mathcal L}_{09}(\mathbf{Q})$ in the $Q_{y}-Q_{z}$ plane have different symmetry under reflection over the $Q_{z}$ axis.
Thus, symmetry reduction occurs by oscillation between these two $\text{Re}\big[{\mathcal L}_{09}(\mathbf{Q})\big]$ and $\text{Im}\big[{\mathcal L}_{09}(\mathbf{Q})\big]$ terms.

Let us analyze the signal at $30\tau$, which deserves a separate discussion. 
Signal depletion is observed in the $Q_{z} = Q_{y}$ direction at both delay times as documented in the 
top panels of Fig.\,\ref{fig03}.
In real space, this direction includes the nitrogen atom that buckles {\it above} the molecular ($yz$) plane.
Regions of signal enhancements (in blue and cyan) are found along the $Q_{z} = -Q_{y}$ direction, coinciding with the $z = -y$ direction, 
which contains the nitrogen atom  buckling {\it below} the molecular ($yz$) plane.
Left-right reflection about the $y=0$ line is the 
only symmetry element remaining in the projection of the molecule in the $yz$ plane, 
which gives rise to the pattern observed in the TRXD signal.
Hence, interference effects due to electronic coherences lead to the symmetry reduction of 
 the TRXD signal.
The asymmetry of the maxima with respect to  
the $Q_z=0$ line observed at both delay times, 10$\tau$ and 30$\tau$,
correlates in both cases with signal depletion for the nitrogen above the corrole plane
or with signal enhancement for the nitrogen below the plane.

\begin{figure}[h!]
\includegraphics[width = 1.0 \linewidth]{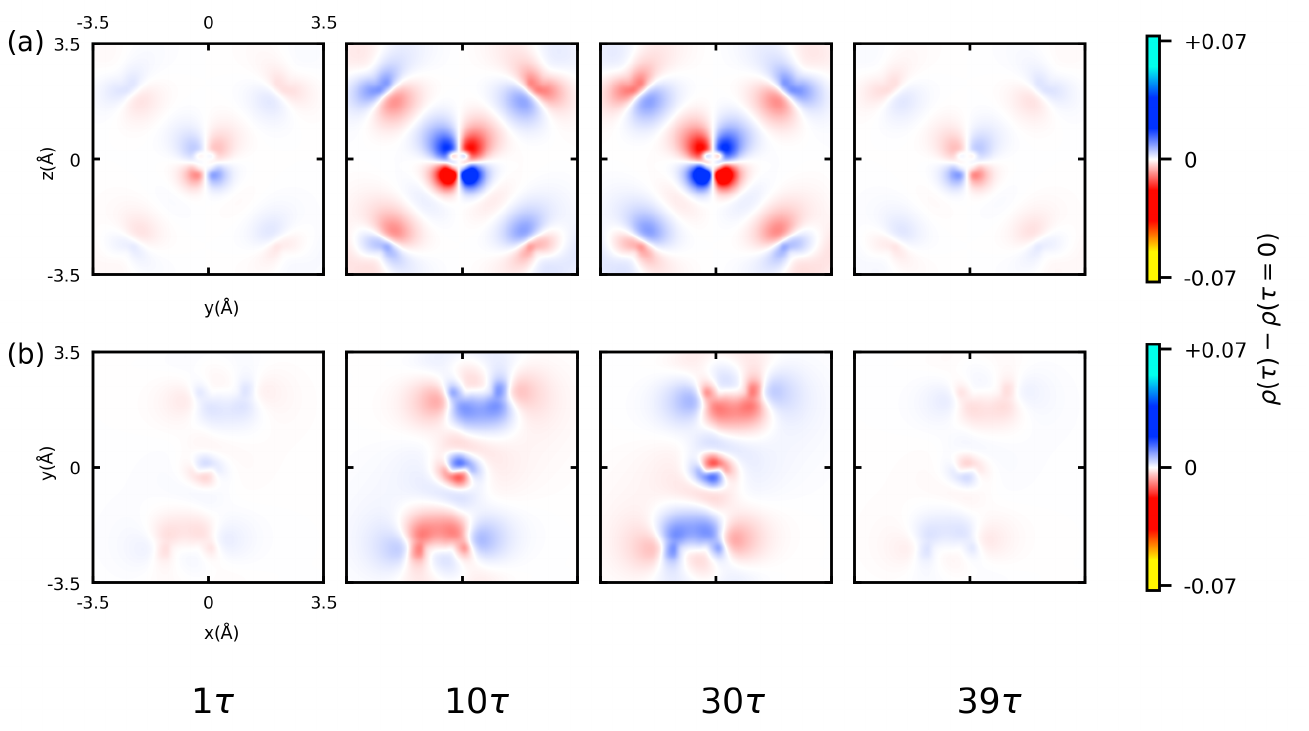}
\caption{Time evolution of the electronic  density of the wave packet 
during field-free charge migration in copper corrole.
The electronic density at zero time delay is subtracted at all subsequent delay 
times and the difference density is represented
in the (a)  $yz$  and (b) $xy$ planes.
As in Fig.\,\ref{fig03},  $\tau = \frac{\textrm{T}}{40}$ is chosen with $\textrm{T} = 1.2$ fs as  the characteristic timescale of the electron dynamics.} \label{fig04}
\end{figure}

The signal in the $Q_{y}-Q_{z}$ plane at the beginning and end of the  characteristic timescale transforms 
according to the B\textsubscript{1} irreducible representation (IRREP) of the C\textsubscript{2v} point group. 
However, at other delay times, the  signal transforms according to the A'' IRREP 
of the C\textsubscript{s} point group [see the second and third figures 
of  Fig.\,\ref{fig03}(a)].
Incidentally, this is also the point group of the molecule projected in the $yz$ plane;  see right panel of Fig.\,\ref{fig01}. 
Let us understand this symmetry alteration  by analyzing 
the time-dependent difference density during field-free charge migration, as shown in
Fig.\,\ref{fig04}. 
As reflected from the top panels, the difference density in the $yz$ plane
is found to transform according to the A'' IRREP of the C\textsubscript{s} point group. This IRREP belongs to
the point group of the plane projection of the molecule. 
Also,  it is known that the Fourier transform of a transition density belonging to the C\textsubscript{s} point group
corresponds to the C\textsubscript{2v} point group~\cite{Defranceschi_1990}.

The temporal evolution of the signals in the $Q_{x}-Q_{y}$ plane for different delay times 
is significantly different than the signals in the $Q_{y}-Q_{z}$ plane [see Figs.~\ref{fig03}(a) and \ref{fig03}(b)]. 
The presence of the twisted structures in the diffraction signals  seems a signature   
of the saddled structure of copper corrole during charge migration. 
Both the TRXD signal [Fig.\,\ref{fig03}(b)] and the difference density [Fig.\,\ref{fig04}(b)] 
retain exactly the same structure at all times, 
albeit with phase reversal and intensity variations during the dynamics.
Both the signal and the difference density 
transform according to the A\textsubscript{u} IRREP of the C\textsubscript{i} point group,  to which
belongs the projection of the molecule in the $xy$ plane. 
It was shown that the Fourier transform of a transition density 
belonging to the C\textsubscript{i} point group will belong to the same group~\cite{Defranceschi_1990}.

To better understand  the connection between the results shown in Figs.~\ref{fig03} and \ref{fig04}, let us 
analyze the key expression for TRXD, which reveals that the difference diffraction signal  encodes the Fourier transform of the transition electron density $\int \mathrm{d}\mathbf{r} ~\langle \Phi_{j}  | \hat{\rho}(\mathbf{r}) | \Phi_{k}  \rangle~ e^{-i \mathbf{Q} \cdot \mathbf{r}}$ [see Eq.~\eqref{eq1}]. 
Also, the difference density of an electronic wave packet consists of several terms of transition electron density,  
i.e., $\langle \Phi_{j}  | \hat{\rho}(\mathbf{r}) | \Phi_{k}  \rangle$.  
Note that it is  well established that the TRXD is  simply not related to the 
Fourier transform of the instantaneous electron density of the wave packet, 
$\int \mathrm{d}\mathbf{r} ~\langle \Psi (\tau)  | \hat{\rho}(\mathbf{r}) | \Psi (\tau)  \rangle~ e^{-i \mathbf{Q} \cdot \mathbf{r}}$; and 
electronic coherences and the transition electron density  play a crucial role in TRXD~ \cite{dixit2012}. 
From the top panels of Fig.~\ref{fig03}, it is evident that the electronic coherences do not destroy
the symmetry relations between the transition density and its Fourier transform at early and later delay  times, but they  induce symmetry reduction in the TRXD signal at intermediate delay times during the dynamics.
This finding is specific to this particular example and the choice of exciting field is likely to affect the times at which  symmetry reduction occurs due to interferences.

\begin{figure}[h!]
\includegraphics[width= 1.0 \linewidth]{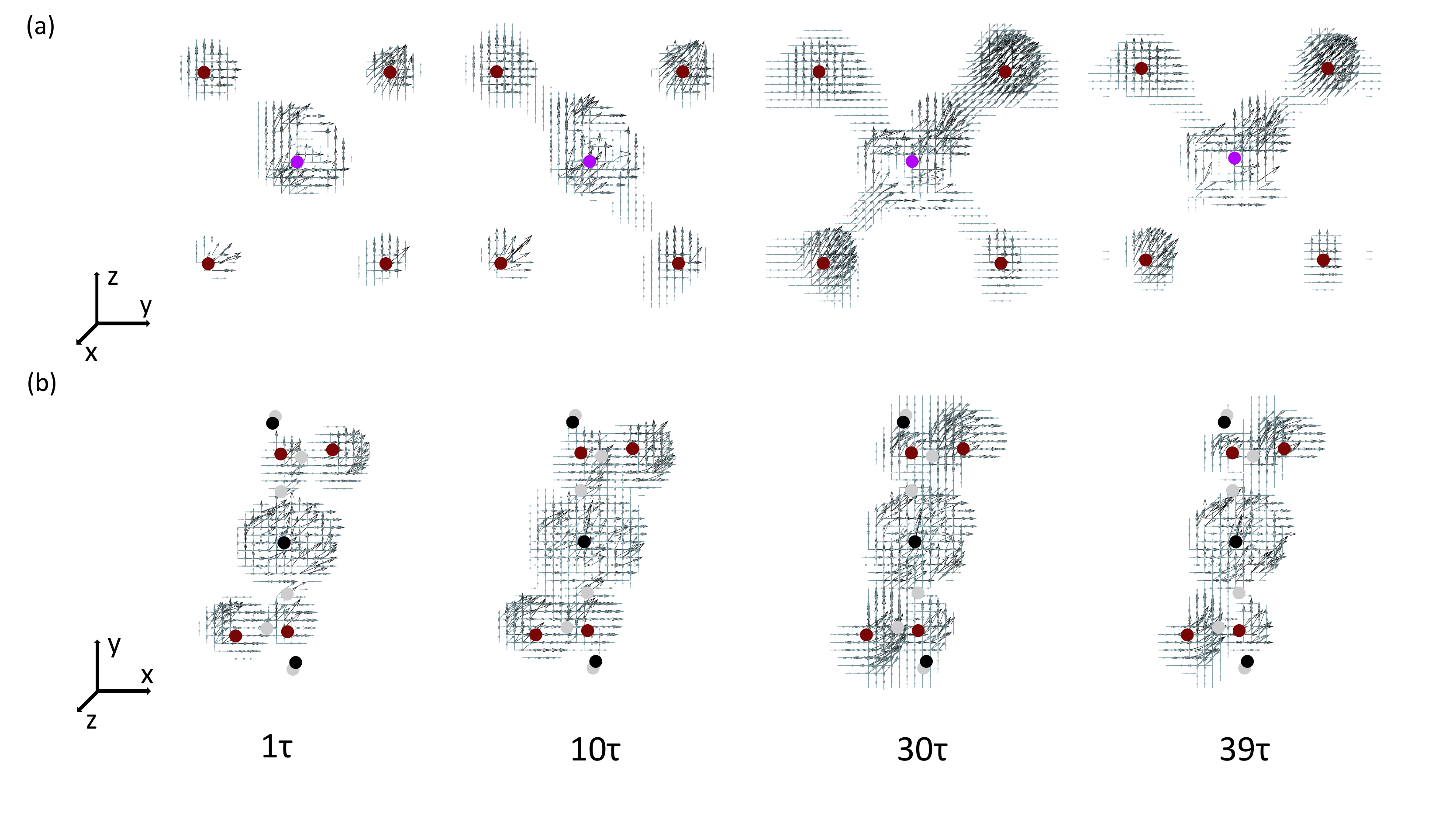}
\caption{Time-dependent electronic flux densities for copper corrole in the
(a) $yz$ and, (b) $xy$ planes at different pump-probe delay times during field-free charge migration.
The laser excitation parameters are the same as in Fig.\,\ref{fig02}.
The period $\tau = \frac{\textrm{T}}{40}$ is chosen as the oscillation period $\textrm{T} = 1.2$ fs of the electron dynamics.
The black, violet, maroon, and gray dots  represent carbon, copper, nitrogen, and hydrogen atoms, respectively.} \label{fig05}
\end{figure}

Let us explore the coherent electron dynamics  in real space to complement the mechanistic picture of the 
charge migration, which could provide  a deeper understanding of the time-resolved diffraction signal in detail. 
Knowledge of the dominant components of the electronic wave packet allows one to 
analyze the electron dynamics at different instances in real space.
Recently, it has been discussed that the analysis  of the transient electronic flux density provides a detailed 
understanding  of the electron dynamics as the  flux density maps the
direction of the electron flow in real space~\cite{hermann2020probing, carrascosa2021mapping, tremblay2021time}.

Figure~\ref{fig05} presents the electronic flux densities associated with the laser-induced charge migration  
at the same instances as in Fig.~\ref{fig03}. 
The central part of the corrole ring containing the nitrogens and the copper is zoomed in
to emphasize  the dominant contribution to the flux densities.  
It is evident from Fig.~\ref{fig05}(a)  that 
the charge migration  is taking place between the nitrogen atoms via the copper atom. 
Most of the charge seems to be displaced from the in-plane nitrogen in the bottom left to the nitrogen atom saddled
above the surface in the top right, which corresponds to the $Q_z = Q_y$ line in momentum space.  
As reflected from Fig.\,\ref{fig03}, the TRXD signals decrease along the $Q_z = Q_y$ line, so 
we assign this migration pattern to a hole displacement during the dynamics.
Synchronously, the nitrogen atom below the plane (top left) and the one in the bottom right appear to feed electrons to the copper atom,
which corresponds well to the region of the TRXD signal increase along the $Q_z = -Q_y$ line in the TRXD signal;  see Fig.~\ref{fig03}(a). 
The view of the flux densities in the $xy$ plane  provides complementary information. 
As evident from Fig.~\ref{fig05}(b), out-of-plane nitrogen atoms are also connected to each other via the copper atom.
This is revealed by the synchronous changes in the direction of the rotation of the  
flux densities around the nitrogen atoms.
At 1$\tau$ and 10$\tau$, the flux densities in the upper two nitrogen atoms rotate anti-clockwise, 
whereas they rotate clockwise for the bottom ones. 
This picture is reversed in the last two time steps at 30$\tau$ and 39$\tau$.

\begin{figure}[h!]
\includegraphics[width= 1.0 \linewidth]{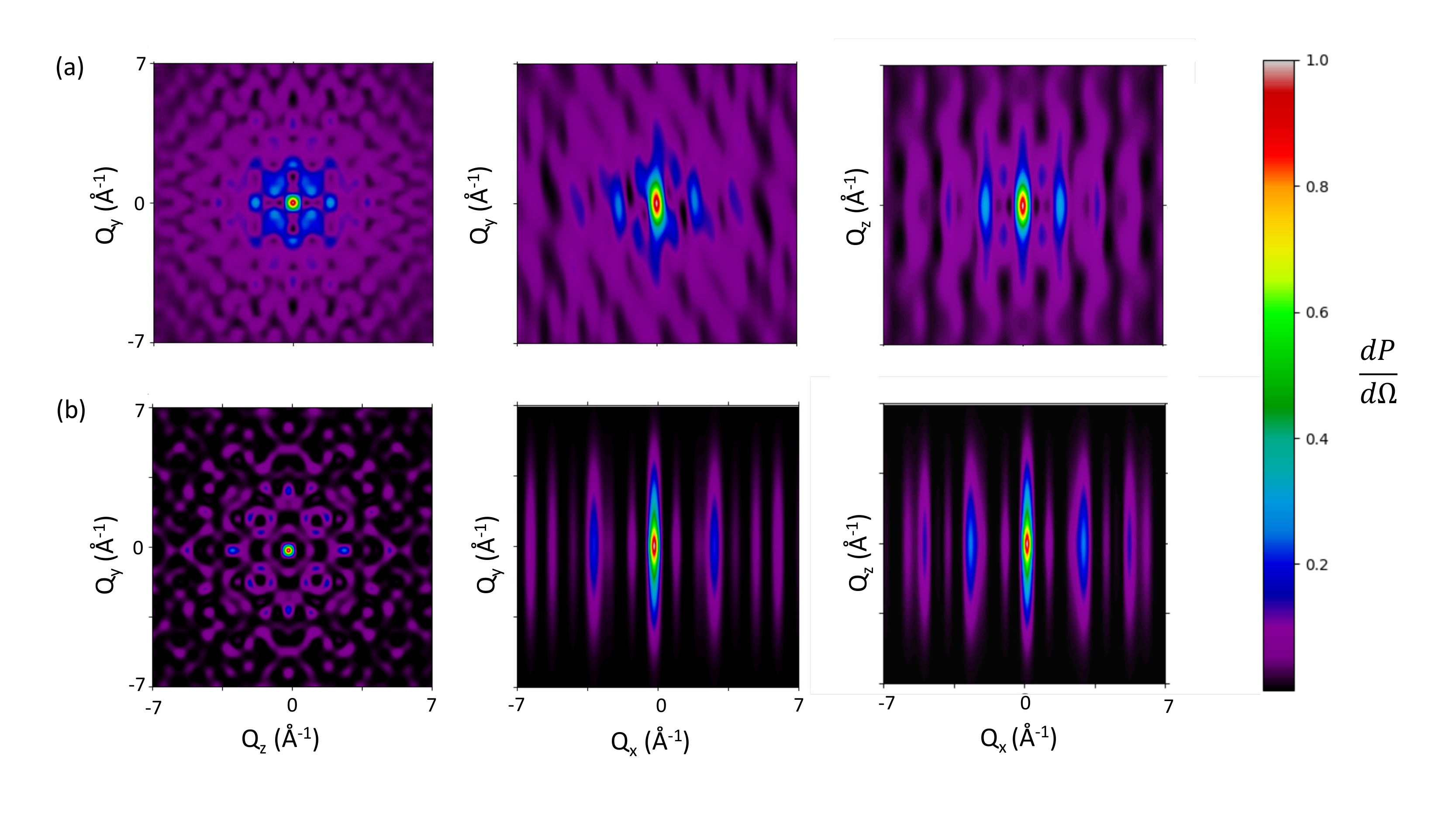}
\caption{The static diffraction signals corresponding to  the ground state for (a) copper corrole and (b) copper
porphyrin in the $Q_{y}-Q_{z}$, $Q_{y}-Q_{x}$ and $Q_{z}-Q_{x}$ planes (from left to right). 
The signals are normalized with respect to their maximum values.} \label{fig06}
\end{figure}

Analysis of the electronic flux densities in  Fig.~\ref{fig05} also shows that 
the flux densities in the diagonal nitrogen atoms are different in magnitudes and directions. 
The present findings are in stark contrast with similar laser-induced charge migration dynamics in planar molecules,  such as benzene and porphyrin in which 
the flux densities and the charge migration dynamics
are observed to be symmetric at all times~\cite{hermann2020probing, barth2006periodic, barth2006unidirectional}.
However, the laser-induced dynamics in a chiral molecule was also studied, 
which reveals that such molecules without any symmetry also exhibit asymmetric 
current transport  ~\cite{giri2021imaging, giri2020time}. 
In this sense, the asymmetry in the electronic flux densities appears to be a signature of the saddling in copper corrole during charge migration dynamics,
which could be measured as a symmetry reduction in the time-resolved diffraction signals.
Although the distortion of the molecular structure due to the saddling occurs in the $xy$-plane, 
the signature of this symmetry reduction would rather be observed in the projection in the plane of the molecule, i.e., the $yz$ plane.
On the other hand, the connection between nitrogen atoms diametrically opposite of the copper atom  
is present in both the real-space and momentum-space views of the electron dynamics. 

To further confirm our claim that the diagonal symmetry is related to the saddled 
structure of the copper corrole, we simulate the static diffraction signal of copper porphyrin in 
the ground state, which exhibits a non-saddled structure.
For copper porphyrin in the ground state, the static diffraction signals are perfectly symmetric along 
the $Q_x$ = 0 and $Q_y$ = 0 planes. Moreover, owing to the non-saddled planar structure of copper porphyrin,
the static diffraction signals in the $Q_{x}-Q_{y}$ and $Q_{x}-Q_{z}$ planes are identical, as reflected 
in Fig.~\ref{fig06}(b). This is not the case for the signal of the saddled, 
nonplanar copper corrole shown in Fig.~\ref{fig06}(a).

\section{Conclusion}  
In summary, the present work is the first step towards understanding the interplay between
ultrafast electron dynamics, structural deformation, and symmetry reduction in  time-resolved x-ray imaging.
We investigate laser-induced dynamics in copper corrole, which has an interesting saddled geometry with only slightly reduced symmetry.
A linearly-polarized pump pulse is used to trigger electron dynamics, which is imaged by TRXD with atomic-scale spatiotemporal resolution. 
We find that the difference diffraction signals are sensitive to the saddled structure, and 
this asymmetry is reflected in the electronic flux densities. 
For the studied excitation, the saddled nitrogen atoms in copper corrole are found to
facilitate the coherent charge migration between nitrogen and copper atoms.
We believe that our results on imaging electron dynamics in symmetry-reduced systems will motivate further
theoretical and experimental research, in particular on light-induced ultrafast processes in various metal-corroles.

\section*{Acknowledgements}
S. G. acknowledges the Council of Scientific and Industrial Research (CSIR) 
for a senior research fellowship (SRF). 
G. D. acknowledges support from the Science and Engineering Research Board (SERB) India 
(Project No. ECR/2017/001460) and the Ramanujan Fellowship (Grant No. SB/S2/ RJN-152/2015).


\end{document}